\documentclass[twocolumn,pra,aps]{revtex4}
\usepackage{graphicx}

\begin{document}

\title{Comment on ``Quantum theory cannot be extended'' \cite{core10}:\\ Falsification of non-covariant extensions requires the before-before experiment.}

\author{Antoine Suarez}
\address{Center for Quantum Philosophy, P.O. Box 304, CH-8044 Zurich, Switzerland\\
suarez@leman.ch, www.quantumphil.org}

\date{July 3, 2010}

\begin{abstract}
The Suarez-Scarani model is a non-covariant (frame-dependent) relativistic (time-ordered causal) nonlocal extension of quantum theory, which is not refuted by the experiments proposed in Reference \cite{core10}. The covariant extensions considered in \cite{core10} are actually self-contradictory and do not require experimental falsification.

\end{abstract}


\maketitle


Bell type experiments demonstrate (within the limits of a few rather eccentric loopholes) correlations which cannot be explained by means of influences propagating at velocity $V \leq c$ \cite{jb}. Thereby these experiments rule out any \emph{local} extension of quantum theory that provides us with additional information about the outcomes of future measurements.

In the paper ``Quantum theory cannot be extended'' Roger Colbeck and Renato Renner claim to go beyond Bell and rule out nonlocal extensions as well. They make three assumptions:

\emph{ST:} Measurements can be associated with well-definite regions in relativistic spacetime.

\emph{QM:} Predictions made by the existing quantum theory are correct.

\emph{FR:} Any measurement settings can be chosen freely. More precisely, they can be chosen such that they are uncorrelated with anything outside of their future lightcone or, equivalently, anything in their past in any reference frame.

On the basis of these axioms Colbeck-Renner present a theorem stating that there cannot exist any extension of quantum theory that provides us with any additional information.\cite{core10}

In this letter I show that the Suarez-Scarani model is a non-covariant (frame-dependent) extension of quantum theory that is not refuted by the Colbeck-Renner theorem. Additionally I argue that the covariant extensions considered in \cite{core10} are actually self-contradictory.

\begin{figure}[t]
\includegraphics[width=80 mm]{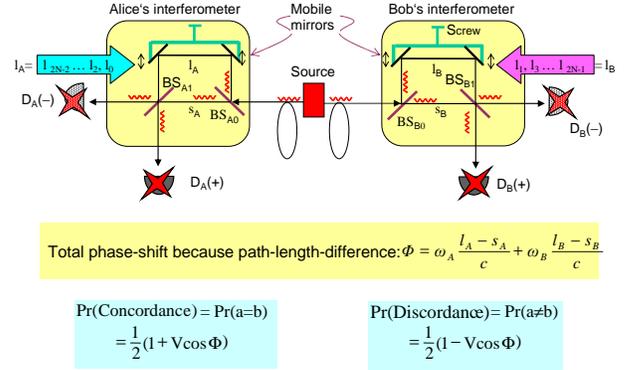}
\caption{Diagram of a chained Bell experiment using interferometers: The source emits photon pairs. Photon A (frequency $\omega_{A}$) enters Alice's interferometer to the left through the beam-splitter BS$_{A0}$ and gets detected after leaving the beam-splitter BS$_{A1}$, and photon B (frequency $\omega_{B}$) enters Bob's interferometer to the right through the beam-splitter BS$_{B0}$ and gets detected after leaving the beam-splitter BS$_{B1}$. The detectors are denoted D$_{A}(\pm)$ and D$_{B}(\pm)$, and correspondingly we say that the detections give the values ($a, b\in\{+1,-1\}$). Each interferometer consists in a long arm of length $l_{i}$, and a short one of length $s_{i}$, $i\in\{A,B\}$. Frequency bandwidths and path alignments are chosen so that only the coincidence detections corresponding to the path pairs: $(s_{A},s_{B})$ and $(l_{A},l_{B})$ contribute constructively to the correlated outcomes in regions A and B, where $(s_{A},s_{B})$ denotes the pair of the two short arms, and $(l_{A},l_{B})$ the pair of the two long arms.
Chained Bell experiments use $N$ different values of $l_{A}$ ($l_0, l_2,...,l_{2N-2}$) and $N$ values of $l_{B}$ ($l_1, l_3,...,l_{2N-1}$), with $N\geq2$.  $\Phi$ is the phase parameter depending on settings $l_A, l_B$ on both sides of the setup.}
\label{f1}
\end{figure}

Consider a Bell experiment (like the sketched in Figure \ref{f1}): A source emits pairs of photons in a entangled state. One of the photons is sent to Alice's laboratory and measured with an interferometer, and the other photon is sent to Bob's laboratory and measured with a similar device. Alice sets the parameter $l_A$ (likewise Bob for $l_B$) and gets the outcome \emph{a} (respectively \emph{b}). Alice's and Bob's measurements happen spacelike separated from each other.

By setting the measuring devices BS$_{A1}$ and BS$_{B1}$ in movement one gets different relativistic inertial frames. We denote:

$t_{a(\underline{A})}$ the time at which Alice's outcome is chosen in the inertial frame of BS$_{A1}$ (and similarly for $t_{b(\underline{A})}$).

$t_{b(\underline{B})}$ the time at which Bob's outcome is chosen in the inertial frame of BS$_{B1}$ (and similarly for $t_{a(\underline{B})})$.

The Suarez-Scarani model \cite{asvs97, as09} states that:

If $t_{a(\underline{A})}<t_{b(\underline{A})}$ the outcome \emph{a} is given by a function depending on local variables: the setting $l_A$ and the variable \emph{u} representing the information accessible from the \emph{past light cone} at the moment of Alice's measurement:
\begin{eqnarray}\label{1}
  a=F_{\underline{A}B}(l_A,u)
\end{eqnarray}
where the subscript $\underline{A}B$ means that Alice's outcome is chosen before Bob's one in the inertial frame of Alice's beam-splitter.

If $t_{a(\underline{A})} \geq t_{b(\underline{A})}$, \emph{a} is given by a function depending on the settings $l_A,l_B$ and the nonlocal variable $\alpha$:
\begin{eqnarray}\label{2}
  a=F_{B\underline{A}}(l_A,l_B,\alpha)
\end{eqnarray}
where the subscript $B\underline{A}$ means that Alice's outcome is chosen after or simultaneously to Bob's one in the inertial frame of Alice's beam-splitter, and $\alpha$ represents the information accessible from the corresponding \emph{past half space} at the moment of Alice's measurement in said frame, and hence includes the local variables $u$ and $v$ as well.

And similarly for Bob's measurement:

If $t_{b(\underline{B})}<t_{a(\underline{B})}$ the outcome $b$ is given by:
\begin{eqnarray}\label{3}
  b=F_{\underline{B}A}(l_B,v)
\end{eqnarray}
where the subscript $\underline{B}A$ means that Bob's outcome is chosen before Alice's one in the inertial frame of Bob's beam-splitter.

And if $t_{b(\underline{B})} \geq t_{a(\underline{B})}$, $b$ is given by:
\begin{eqnarray}\label{4}
  b=F_{A\underline{B}}(l_A,l_B,\beta)
\end{eqnarray}
where the subscript $A\underline{B}$ means that Bob's outcome is chosen after or simultaneously to Alice's one in the inertial frame of Bob's beam-splitter, and $\beta$ represents the information accessible from the corresponding \emph{past half space} at the moment of Bob's measurement in said frame, and hence includes the local variables $u$ and $v$ as well.

The functions expressed by the equations (\ref{2}) and (\ref{4}) are supposed to generate joint probabilities $P(a,b)$ reproducing the quantum mechanical correlations. With before-before timing (each beam-splitter selecting before in the own reference frame) the functions in (\ref{1}) and (\ref{3}) generate outcomes such that the joint probability $P(a,b)$ fulfills the Bell inequalities but it is not necessarily product ($P(a)\times P(b)$) \cite{as09}.

To date there is no proof that correlations defined according to (\ref{1}), (\ref{2}), (\ref{3}) and (\ref{4}) are signaling \cite{vang}, and one can conjecture that the quantum mechanical formalism prevents such a proof.

The Suarez-Scarani model combines local and nonlocal hidden variables to a full time-ordered causal nonlocal explanation of the quantum correlations, which remains compatible with any of the experiments having confirmed relativity theory, and Bell-type experiments as well. For the time being it cannot be excluded as signaling. While the model can be considered relativistic, because it accepts multisimultaneity, it is not Lorentzinvariant, since it assumes frame-dependent distributions. It illustrates that Lorentzinvariance is not a necessary condition for non-signaling \cite{as10}.

So far the model is compatible with the axioms \emph{ST} and \emph{FR} in \cite{core10}, but in before-before experiments it makes predictions conflicting with quantum mechanics: If each measuring device, in its own reference frame, is first to select the output, the nonlocal dependencies (\ref{2}) and (\ref{4}) become irrelevant and only (\ref{1}) and (\ref{3}) matter. Hence the model predicts disappearance of nonlocal correlations with maintenance of possible local ones \cite{as09}. By contrast quantum mechanics predicts timing independent nonlocal correlations.

In their argument Colbeck-Renner exclude probability distributions depending on the time order (see Footnotes 3 and 6 in \cite{core10}) for no apparent reason, as far as I can see. Certainly, covariance implies: $F_{\underline{A}B}=F_{B\underline{A}}$ and $F_{\underline{B}A}=F_{A\underline{B}}$ \cite{ng}. However, as stated above, it is possible violating covariance while respecting multisimultaneity and non-signaling \cite{as10}.

In addition to time-order independent (covariant) nonlocal influences quantum theory makes the assumption of random uniform distributed outcomes. Since covariant nonlocal ``hidden'' variables are impossible \cite{ng}, the only way for covariant extensions to provide additional information about future measurements is through biased random outcomes $a$ and $b$. The bias can come from some nonlocal system and not necessarily from a local hidden variable. It is this kind of covariant extensions (more general than those adding ``a classical list assigning outcomes''), which is specifically addressed by the Colbeck-Renner theorem \cite{core10}.

Accordingly, if one assumes dependencies (\ref{1}) and (\ref{3}) generating random uniform distributed local outcomes, the corresponding Suarez-Scarani model provides a non-covariant (frame-dependent) full deterministic extension of quantum theory which is not tested by the proposed Colbeck-Renner experiments \cite{core10} (and even less by Leggett-type experiments \cite{gro,cb}). To test the extension one needs before-before experiments with beam-splitters in motion \cite{as09}.

Before-before experiments have been done, and falsify time-ordered nonlocal influences (i.e., \emph{nonlocal determinism}) \cite{szsg}. Since these experiments rule out multisimultaneity and demonstrate that ``the space-time does not contain the whole physical reality'' (Nicolas Gisin) \cite{as10}, they can be considered to falsify relativity while confirming covariance. Nonlocal extensions of quantum mechanics (like Colbeck-Renner and Leggett ones), as far as they assume nonlocal determinism, are falsified by the before-before experiment as well \cite{as09}.

Conversely, the before-before experiment does not suffice to rule out covariant extensions that give some information about the outcomes without determining them completely, for instance, by adding subensembles of states leading to biased joint outcomes (and thereby biased local ones as well). I consider now such covariant extensions proposed for experimental verification in \cite{core10}, and prove that they are self-contradictory.

Consider the chained Bell experiment sketched in Figure \ref{f1}. Suppose one of the measurements produces the value $a$ ($a\in\{+1,-1\}$), and the other the value $b$ ($b\in\{+1,-1\}$). In agreement with the available observations (but independently of the quantum mechanical description) one can assume that the probability $P(a, b)$ of getting the joint outcome $(a, b)$ depends on the choice of the phase parameter $\Phi(l_A, l_B)$, and introduce the following conditional probabilities:

\begin{eqnarray}\label{5}
    P(a=b|\Phi(l_A, l_B)) \nonumber\\
    P(a\neq b|\Phi(l_A, l_B))
\end{eqnarray}
where the probabilities in (\ref{5}) do not depend on the path lengths ($l_A, l_B$), if these yield the same value of $\Phi$.

\begin{figure}[t]
\includegraphics[width=80 mm]{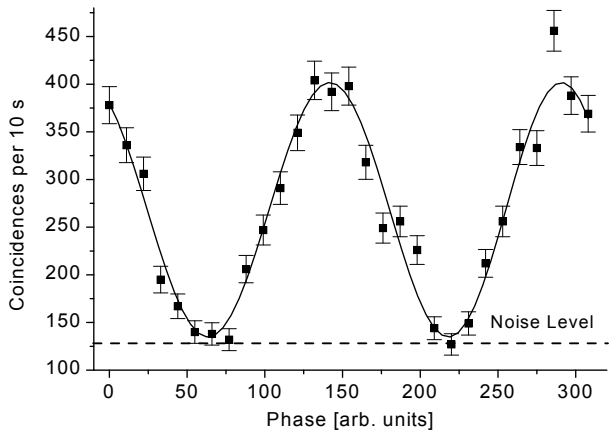}
\caption {Experimental data gathered in \cite{szsg}: ``Coincidence counts per 10 s'' measures $P(a=b=1|\Phi(l_A, l_B))$ for different phase values $\Phi(l_A, l_B)$. The dashed line indicates the noise level. The visibility $V$ after subtraction of the noise is $97\pm5\%$.}
\label{f2}
\end{figure}

In different entanglement experiments, and in particular in \cite{szsg}, one measures accurately the probabilities $P(a=b|\Phi)$ over several periods of $\Phi$ by changing the arm length in only one of the interferometers. Figure \ref{f2} represents measurements of $P(a=b=1|\Phi)$ for different phase values $\Phi$ in the experiment described in \cite{szsg}.

From these measurements (Figure \ref{f2}) one knows that the probability $P(a, b)$ of getting the joint outcome $(a, b)$ is well described by the following equations:
\begin{eqnarray}\label{6}
    P(a=b|\Phi)=\frac{1}{2}(1+V\cos\Phi) \nonumber\\
    P(a\neq b|\Phi)=\frac{1}{2}(1-V\cos\Phi)
\end{eqnarray}
\noindent where $\Phi=\omega_{A}(l_A - s_A)/c+\omega_{B}(l_B - s_B)/c$, and $V$ is a visibility factor depending mainly on the efficiency of the detectors. The data of Figure \ref{f2} were obtained with $V=0.97$. The Equations \ref{6} are an extrapolation of these data for other values of the visibility. The extrapolation can be easily verified as soon as the corresponding $V$ is experimentally achievable.

Consider now the function $I(N)$ defined as:
\begin{eqnarray}\label{7}
    I(N)&=& P(a=b|\Phi(l_{0},l_{2N-1})) \nonumber\\
    &+& P(a\neq b|\Phi(l_{0},l_{1})) \nonumber\\
    &+& P(a\neq b|\Phi(l_{1},l_{2})) \nonumber\\
    &+& ....... \nonumber\\
    &+& P(a\neq b|\Phi(l_{2N-2},l_{2N-1}))
\end{eqnarray}
\noindent where $P(a\!=\!b|\Phi(l_{0},l_{2N-1}))$ means the conditional probability that Alice and Bob get the same outcome if the phase's value results from long interferometers' arms set to $l_{0},l_{2N-1}$, and $P(a\!\neq\! b|\Phi(l_{i},l_{i+1}))$ the conditional probability that Alice and Bob get different outcomes if the phase's value results from long interferometers' arms set to $l_{i},l_{i+1}$; depending on $i$, $l_{i}$ denotes the arm of Alice's or Bob's interferometer.

For each $N$, $I(N \geq1$ defines a Bell inequality or locality criterion. $I(2)\geq1$ represents the well known CHSH inequality for experiments with 4 measurements. Accordingly, $I(N)<1$ defines correlations that cannot be explained by means of local influences. If $I(N)$ is interpreted as a measure of nonlocality the ``maximum'' is reached for $I(N)=0$ \cite{bkp}.

For convenience we assume that any two values $l_{i}, l_{i+1}$, with $i\in\{0, 2N-2\}$, in (\ref{7}) define the same phase parameter, resulting from the equipartition of a value $\Theta$:
\begin{equation}\label{8}
\Phi(l_{i}, l_{i+1})=\Theta/2N
\end{equation}

By substituting (\ref{8}) into equation (\ref{7}) one obtains:
\begin{eqnarray}\label{9}
    I(N,\Theta)&=& P\left(a=b\left|(2N-1)\frac{\Theta}{2N}\right.\right)\nonumber\\
    &+& (2N-1)P\left(a\neq b\left|\frac{\Theta}{2N}\right.\right)
\end{eqnarray}
\noindent where now we use the notation $I(N,\Theta)$ to indicate that $I(N)$ depends on the variable $\Theta$ as well.

Consider now chained Bell experiments with $N$ settings at each side chosen at random (Figure \ref{f1}). Nonlocal extensions can't help acknowledging the experimental data of Figure \ref{f2}. Additionally, since they take nonlocality for granted, they take for granted that the probabilities (\ref{6}) also hold if the measurement is picked at random from those specified in (\ref{7}).

Substituting (\ref{6}) into (\ref{9}) one is led to:
\begin{eqnarray}\label{10}
    I(N,\Theta)&=& \frac{1}{2}\left(1+Vcos\left(\Theta-\frac{\Theta}{2N}\right)\right)\nonumber\\
    &+& \frac{2N-1}{2}\left(1-Vcos\frac{\Theta}{2N}\right)
\end{eqnarray}

For $\Theta=\pi$ equation (\ref{10}) gives:
\begin{equation}\label{11}
    I(N,\pi)=N(1-V cos \frac{\pi}{2N})
\end{equation}

\begin{figure}[t]
\includegraphics[width=80 mm]{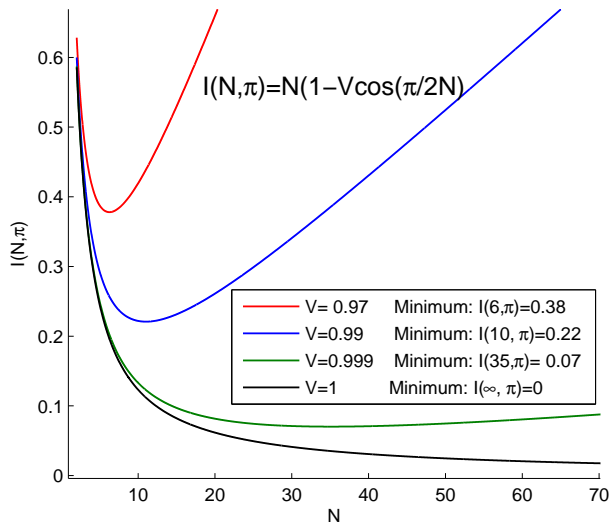}
\caption{The function $I(N,\pi)$ for different values of the visibility $V$. $V=1$ corresponds to the theoretical quantum mechanical prediction (see text).}
\label{f3}
\end{figure}

Figure \ref{f3} represents the function (\ref{11}) and gives its minimum for different values of the visibility. $V=1$ is the theoretical prediction of quantum mechanics.

Notice that the minima in Figure \ref{f3} are obtained from data, and in these sense are theory independent.

We denote $D$ the statistical distance between the distribution of the outcomes predicted by the extension and the uniform random distribution of quantum theory. In case of Leggett-type extensions tested to date $D$ measures a dependence on a local hidden polarization \cite{cb,gro}. However $D$ can come from some general system, not necessarily local hidden variables.

As proved in \cite{core10}, the non-signaling condition implies:
\begin{equation}\label{12}
D\leq \frac{3I(N)}{2}
\end{equation}

Suppose an extension assuming a variational distance of $D = 0.25$. According to  Figure \ref{f3} the same extension predicts for visibility $V=0.999$ the minimum $I(35, \pi)=0.07$, and Equation (\ref{12}) gives  $D\leq \frac{3}{2}\; 0.07=0.11$. Therefore, the extension is self-contradictory. For any possible distance one can find a minimum leading to a contradiction.\\
\indent Notice however that an extension assuming a variational distance $D$ depending on $\Theta/2N$ always fulfills (\ref{12}), and is neither falsified by this argument nor by whatever chained Bell experiment \cite{as09a}.\\
\indent In conclusion, any covariant extension has to match the experimental results of Figure \ref{f3} and fulfill the Colbeck-Renner inequality (\ref{12}). These two requirements exclude extensions with variational distance $D$ that is independent of $N$. Consequently, covariant extensions (with $N$-independent $D$) can be considered falsified without need of experiments different than the conventional Bell ones (N=2). This holds in particular for Leggett-type models \cite{cb, gro}.\\
\indent Nonetheless chained Bell experiments with $N>2$ and settings chosen at random may be useful to the aim of building a cryptographic tool allowing us to upper bound the local bias.\\
\indent According to these conclusions it seems that only frame-dependent (non-covariant) extensions of quantum theory require trials of a type different than the conventional Bell experiments. This stresses the interest of deciding whether the Suarez-Scarani model is signaling.\\

\noindent \emph{Acknowledgments}: I am grateful to Cyril Branciard, Roger Colbeck, Nicolas Gisin, Stefano Pironio, and Renato Renner for insightful discussions.

\end{document}